# Broken Symmetry States and Divergent Resistance in Suspended Bilayer Graphene

Benjamin E. Feldman, Jens Martin, Amir Yacoby*
*Department of Physics, Harvard University, Cambridge, MA 02138, USA*
*e-mail: yacoby@physics.harvard.edu

**Graphene[1] and its bilayer have generated tremendous excitement in the physics community due to their unique electronic properties[2]. The intrinsic physics of these materials, however, is partially masked by disorder, which can arise from various sources such as ripples[3] or charged impurities[4]. Recent improvements in quality have been achieved by suspending graphene flakes[5,6], yielding samples with very high mobilities and little charge inhomogeneity. Here we report the fabrication of suspended bilayer graphene devices with very little disorder. We observe fully developed quantized Hall states at magnetic fields of 0.2 T, as well as broken symmetry states at intermediate filling factors $\nu$ = 0, ±1, ±2 and ±3. The devices exhibit extremely high resistance in the $\nu$ = 0 state that grows with magnetic field and scales as magnetic field divided by temperature. This resistance is predominantly affected by the perpendicular component of the applied field, indicating that the broken symmetry states arise from many-body interactions.**

The linear dispersion of graphene near its Fermi energy gives rise to low-energy excitations that behave as massless Dirac fermions[2]. These quasiparticles exhibit an anomalous integer quantum Hall effect[7,8] in which the Hall conductivity is quantized at values of $\sigma_{xy} = \nu e^2/h$ for filling factors $\nu = 4(N + 1/2)$. Here, $N$ is an integer, $e$ is the electron charge, $h$ is Planck's constant, and the factor of 4 is due to spin and valley degeneracy. Recent measurements[9,10] of graphene monolayers in high magnetic field $B$ have revealed additional broken symmetry quantum Hall states at $\nu$ = 0, ±1 and ±4, which have been proposed to arise due to quantum Hall ferromagnetism (QHF)[11-13] or the formation of excitonic energy gaps[14,15]. The $\nu$ = 0 state has received particular attention due to contradictory experimental observations. Some samples exhibit large magnetoresistance of ~$10^5$-$10^7$ Ω near the charge neutrality point[16-19], and this behavior has been ascribed to the opening of a spin gap[16], the approach of a field-induced

Kosterlitz-Thouless transition to an insulating state[17,18] or the formation of a collective insulator[19]. Others, however, report[9,10,20] resistance of only ~$10^4$ Ω, and attribute their findings to the existence of spin polarized counter-propagating edge modes[9,20].

While experimental investigations of broken symmetry quantum Hall states have so far focused only on graphene monolayers, recent theoretical studies have investigated QHF in bilayer graphene[21] and the resultant ground states at intermediate filling factors[22]. The physics is richer in bilayers due to an additional twofold orbital degeneracy in the Landau level (LL) spectrum[23] which leads to an eightfold degenerate LL at zero energy and a corresponding step of $8e^2/h$ in $\sigma_{xy}$. It has been shown both theoretically[24] and experimentally[25,26] that a potential difference between the two layers opens an energy gap, leading to a plateau in $\sigma_{xy}$ at $\nu = 0$, but no other broken symmetry states have been observed. Here we report the fabrication of high quality suspended bilayer graphene devices (Figs. 1a and 1b) that exhibit full splitting of the eightfold degenerate zero-energy LL. The $\nu = 0$ state appears at $B \approx 0.1$ T and is characterized by an extremely large resistance that increases exponentially with the perpendicular component of $B$. The $|\nu| = 2$ states emerge at $B = 0.7$ T, and all symmetries are broken for $B \geq 3$ T.

We focus first on the behavior of our samples in zero magnetic field. Fig. 1c shows the resistivity $\rho$ of two suspended bilayers as a function of carrier density $n$. Each sample displays a sharp peak in $\rho$ with a full width at half maximum on the order of $10^{10}$ $cm^{-2}$, comparable to suspended monolayer devices[5,6] and an order of magnitude smaller than unsuspended bilayers[27]. In all samples, the peak lies close to zero back gate voltage ($|V_{peak}| < 0.5$ V), indicating that there is little extrinsic doping in our devices. As a measure of sample cleanliness, we can estimate the magnitude of carrier density fluctuations $\delta n$ based on the carrier density dependence of the conductivity $\sigma(n)$, shown in Fig. 1d. Near the charge neutrality point, local variations in potential lead to the formation of electron-hole puddles[28], and $\sigma(n)$ is expected[29] to remain constant in this regime because $|n| < \delta n$. In our suspended bilayers, $\delta n$ is typically on the order of $10^{10}$ cm$^{-2}$, and it reaches as low as $10^{9}$ cm$^{-2}$ in sample S3.

The temperature dependence of the minimum conductivity $\sigma_{min}$ (Fig. 1e) provides a second method to estimate $\delta n$. At low temperatures, $\sigma_{min}$ is dominated by transport through the electron-hole puddles rather than thermal effects, so we expect strong temperature dependence only for $k_BT > E_{pud}$, where $k_B$ is Boltzmann's constant and $E_{pud}$ is the typical magnitude of the screened potential fluctuations responsible for electron-hole puddles. For bilayer graphene, we can estimate $E_{pud} \approx h^2\delta n/8\pi m^*$, where $m^* \approx 0.033m_e$ is the effective mass in bilayer graphene[29] ($m_e$ is the electron mass). In sample S3, $\sigma_{min}$ shows temperature dependence down to 450 mK, providing an upper bound of $\delta n < 10^{9}$ cm$^{-2}$. In contrast, $\sigma_{min}$ saturates at approximately 2 K in sample S4, corresponding to $\delta n \approx 5\text{x}10^{9}$ cm$^{-2}$. Both estimates are consistent with the estimate of disorder obtained from $\sigma(n)$. In both samples, $\sigma_{min}$ at 450 mK is a few times the conductance quantum, in good agreement with theoretical predictions for its intrinsic limit[30-32].

In contrast to the typically reported linear behavior in bilayer graphene, $\sigma(n)$ is sublinear in suspended samples (Fig. 1f). If we assume mobility $\mu = (1/e)\mathrm{d}\sigma/\mathrm{d}n$, then $\mu$ typically ranges from 10,000-15,000 cm$^2$/V·s in our suspended bilayers at carrier density $n$ of 2-3x$10^{11}$ cm$^{-2}$. These numbers represent a modest improvement of approximately a factor of two over unsuspended bilayers, but it remains unclear why the mobility is this low given the indications of sample quality discussed above, the low magnetic field at which we observe quantum Hall plateaus, and the high mobilities observed in suspended monolayers[5,6]. Adam and Das Sarma predict[29] that the mobility of bilayer graphene should be more than an order of magnitude smaller than that of monolayer graphene. This discrepancy was not observed in unsuspended samples[27], but mobility in such samples may be limited by disorder associated with the substrate. It is also worthwhile to comment on the possibility that the sharp dip in conductivity at low $n$ is enhanced by a small energy gap that opens due to disorder-induced differences in carrier density between the top and bottom layers of the flake[21]. Differences in density of a few times $10^9$ cm$^{-2}$ would lead to an energy gap[26] of approximately 0.3 meV.

We next discuss the magnetic field dependent behavior of our samples. Figs. 2a and 2b show the conductance of sample S1 as a function of magnetic field and carrier density $G(n, B)$, and Fig. 2c highlights traces of $G(n, B)$ at several representative magnetic fields. Our devices exhibit the expected quantum Hall conductance plateaus at $4me^2/h$ for bilayer graphene, corresponding to filling factors $\nu = \pm 4m$ (black dotted lines in Fig. 2). Full quantization for $\nu = \pm 4$ occurs at very low $B$, indicative of the cleanliness of our devices. For sample S3, the $\nu = \pm 4$ plateaus are fully quantized at 0.2 T (inset of Fig. 2d).

In addition to the expected behavior highlighted above, we observe quantum Hall plateaus corresponding to intermediate filling factors $\nu = 0, \pm 1, \pm 2$ and $\pm 3$ (colored dotted lines in Fig. 2). The $|\nu| = 2$ (1) state becomes apparent at 0.7 (2.7) T, and fully develops into a conductance plateau of $2e^2/h$ ($e^2/h$) at 2.7 (7.3) T on the hole side (Figs. 2a and 2d). The $|\nu| = 3$ state appears at a similar magnetic field to the $|\nu| = 1$ state, but leaves the experimentally accessible regime before it is fully quantized. Near the charge neutrality point, a $\nu = 0$ state with a very large resistance that increases exponentially with $B$ emerges at $B \approx 0.1$ T. Measurements of Hall bar devices show a corresponding plateau at $\sigma_{xy} = 0$ and rule out the possibility that the large resistance arises from contact resistance between the graphene and the electrical leads. We focus, however, on two-terminal devices because they are more homogeneous (see supplementary information).

The appearance of quantum Hall states at $\nu = 0, \pm 1, \pm 2$ and $\pm 3$ indicates that the eightfold degeneracy of the zero-energy LL in bilayer graphene is completely lifted in our samples. The magnetic field at which these effects appear is over an order of magnitude smaller than has been reported for monolayers[9,10,16-20]. Broken symmetry states could arise from multiple causes, including spin splitting due to the Zeeman effect[16], strain-induced lifting of valley degeneracy[33], the opening of an energy gap due to a potential difference between the two layers[24], or QHF[21]. In our samples, the proximity of $V_{peak}$ to zero back gate voltage makes it unlikely that we observe an energy gap due to chemical doping[26]. It has recently been shown[34] that large-scale ripples appear in suspended graphene membranes when they are cooled from 600 to 300 K, but room temperature scanning electron micrographs of our suspended flakes do not show prominent corrugations (Fig. 1a). The interaction energy due to QHF in bilayer graphene is expected to be two orders of magnitude stronger than spin splitting caused by the Zeeman effect[21], so the observed broken symmetry states are unlikely to be associated with Zeeman splitting. We therefore tentatively attribute the symmetry breaking to QHF. The order in which broken symmetry states appear in our samples is indeed consistent with the expectations of Barlas *et al.*, who predict[21] the largest energy gap for a spin polarized state at $\nu = 0$, followed by spin and valley polarized states at $|\nu| = 2$, and finally spin, valley and LL index polarized states at $|\nu| = 1$ and $|\nu| = 3$.

We now discuss in more detail the large magnetoresistance of the $\nu = 0$ state. Fig. 3 shows the maximum resistance of sample S3 in a small carrier density range around the charge neutrality point as a function of magnetic field and temperature $R_{max}(B, T)$ at various temperatures between 450 mK and 24.5 K (See also the supplementary information). $R_{max}(B, T)$ increases by more than four orders of magnitude to $10^8$ Ω (the *de facto* limit of our measurement capabilities) within a few Tesla for $T < 5$ K. This increase is significantly steeper than in monolayers, where the reported[18] resistance reached only 40 MΩ at 30 T. Our data do not fit a Kosterlitz-Thouless type transition, nor do the flakes exhibit activated behavior over the full temperature range of our measurements.

One of the main findings of this report is that $R_{max}(B, T)$ scales as $B/T$, as plotted in Fig. 4a. For $T \geq 1.9$ K, the data collapse quite nicely onto one curve. At lower temperatures, $R_{max}(B, T)$ continues to increase with decreasing $T$, but it does not do so as quickly as expected for $B/T$ dependence (inset of Fig. 4a). This can be explained if we assume that the LLs are broadened by disorder. In such a scenario, a constant offset in magnetic field $B_{off}$ is needed to resolve distinct quantum Hall states. Using $B_{off} = 0.14$ T, in reasonable agreement with the field at which the $|\nu| = 4$ states becomes fully quantized and the $\nu = 0$ resistance begins to diverge (inset of Fig. 4b), the $R_{max}(B, T)$ data collapse onto one curve for the entire temperature range when plotted against $(B - B_{off})/T$ (Fig. 4b).

We infer that an energy gap $\Delta \sim 0.3\text{-}0.9(\mathrm{B[T]})$ meV develops in an applied magnetic field. The gap is several times larger than expected for Zeeman splitting, and tilted field experiments provide further evidence that the broken symmetry states likely arise from QHF rather than Zeeman splitting. $R_{max}(B, T)$ is primarily dictated by the perpendicular component of field $B_{perp}$ (Figs. 4c and 4d), in stark disagreement with the behavior expected for a Zeeman gap. Moreover, at fixed $B_{perp}$, an increase in the parallel component of the field reduces $R_{max}(B, T)$ (Fig. 4d), indicating that the low-energy excitations of the $\nu = 0$ state are not skyrmionic spin flip in nature[22,35]. The linear dependence of $\Delta$ on $B$ is qualitatively different from what is expected for QHF, which predicts[21] a gap that scales as $B^{1/2}$. It is worth noting, however, that early studies[36,37] of the exchange-enhanced spin gap at $\nu = 1$ in GaAs samples also showed an energy gap that was linear in $B$. QHF predicts[21] $\Delta \sim 100$ meV for magnetic fields of a few Tesla, far larger than we observe, but this discrepancy is likely due to disorder in our samples.

**Methods:**

Suspended bilayer graphene devices are fabricated using a method similar to that described[5] by Bolotin *et. al*. Briefly, mechanical exfoliation of highly oriented pyrolytic graphite (Grade ZYA, SPI supplies) is used to deposit few-layer graphene flakes on a Si substrate coated with a 300 nm layer of $SiO_2$. Deposition is carried out at 180 ºC to minimize the amount of water present on the substrate. Bilayer flakes are identified using an optical microscope, based on contrast between the flake and the surrounding substrate. Electrical leads are then patterned using electron beam lithography, followed by thermal evaporation of 3 nm of Cr and 100 nm of Au, and subsequent liftoff in warm acetone. The entire substrate is then immersed in 5:1 buffered oxide etch for 90 s, which etches approximately 40% of the $SiO_2$, including the area under the graphene[5], but not the area under the metal contacts, which extend across the entire width of the flake in order to improve structural integrity. Samples are quickly transferred to methanol and dried using a critical point dryer. Finished samples are transferred to the measurement system as quickly as possible, and are typically used without further cleaning or current annealing. Electronic transport measurements have been performed on multiple samples, using standard AC lock-in techniques with excitation voltages below 100 μV, in either an ultra-high vacuum He-3 cryostat or a dilution refrigerator. The Si substrate serves as a global back gate, which is used to vary the carrier density in the bilayer. Back gate voltage is limited to $|V_{bg}| < 10$ V in order to avoid structural collapse of suspended devices.

**Acknowledgments:**

We would like to acknowledge useful discussions with L. S. Levitov, R. Nandkishore, D. A. Abanin, A. H. Castro Neto, A. H. MacDonald, M. S. Rudner and S. Sachdev. We acknowledge support from Harvard NSEC, Harvard CNS and from the ONR MURI program.

**Competing Financial Interests:**

We have no competing financial interests.

**Figure Legends**

**Figure 1 | Characterization of suspended bilayers at zero magnetic field. a,** False color scanning electron micrograph of a suspended bilayer graphene flake. The scale bar is 1 μm. **b,** Optical microscope image of several two-terminal suspended bilayer samples in series. The scale bar is 1 μm. **c,** Two-terminal resistivity $\rho$ of samples S3 (blue) and S4 (red) as a function of carrier density $n$. Both samples display a pronounced peak in $\rho$ with full width at half maximum of $1.5 \times 10^{10}$ cm$^{-2}$ and $2 \times 10^{10}$ cm$^{-2}$, respectively, at temperature $T = 450$ mK. **d,** Electron and hole branches of the conductivity $\sigma$ for samples S3 (blue) and S4 (red) at $T = 450$ mK. The width of the plateau in $\sigma$, marked by the arrows, indicates the magnitude of carrier density fluctuations due to disorder, estimated to be $10^9$ cm$^{-2}$ in sample S3 and $4 \times 10^9$ cm$^{-2}$ in sample S4. **e,** Temperature dependence of the minimal conductivity $\sigma_{min}$ in samples S3 (blue) and S4 (red). Inset: zoom in on the low-temperature behavior. For sample S4, disorder causes $\sigma_{min}$ to saturate for $T < 2$ K. The decrease of $\sigma_{min}$ for sample S3 down to 450 mK indicates that it is cleaner, consistent with the findings in Figures 1c and 1d. **f,** Conductivity at $T = 450$ mK for samples S3 (blue) and S4 (red). For $n > 2 \times 10^{11}$cm$^{-2}$, the mobility is about 7,500 cm$^2$/V·s. The pronounced dip in the conductivity at very low densities may be enhanced by a disorder-induced gap.

**Figure 2 | Splitting of the eightfold degenerate Landau level in suspended bilayers. a,** Carrier density and magnetic field dependence of the two-terminal conductance $G(n, B)$ in sample S1 at $T = 100$ mK. Lines indicate filling factors $\nu = 8$ and 4 (black), 3 (blue), 2 (purple), 1 (red) and 0 (green). Conversion between back gate voltage and density was calibrated using these filling factors. **b,** 3D rendering of $G(n, B)$ in sample S1. The numbers indicate filling factor. Broken symmetry states at $\nu = 0, \pm 1, \pm 2$, and $\pm 3$ are clearly visible. **c,** Line traces of $G(n,$

$B$) at various magnetic fields. Quantum Hall plateaus associated with the broken symmetry quantum Hall states are apparent. **d,** Conductance traces taken along the dotted lines in Figure 2a. For sample S1, full quantization is observed at $B = 0.4$ T for $\nu = 4$, $B = 2.3$ T for $\nu = 2$, and $B = 7.3$ T for $\nu = 1$. Inset: for sample S3, quantization of the $\nu = 4$ state is reached for $B \approx 0.2$ T at $T = 450$ mK.

**Figure 3 | Temperature and field dependence of the $\nu = 0$ state.** Maximum resistance of sample S3 at the charge neutrality point as a function of magnetic field and temperature. Inset: zoom in on the low-temperature curves. We do not observe saturation of the resistance for temperatures down to 450 mK.

**Figure 4 | Scaling of the maximum resistance in the $\nu = 0$ state. a,** $R_{max}(B, T)$ of sample S3 plotted versus $B/T$. The data collapse onto one curve for temperatures $T > 1.9$ K. Inset: $B/T$ scaling does not succeed for $T < 1.9$ K. **b,** $R_{max}(B, T)$ versus $(B - B_{off})/T$. All data collapse using $B_{off} = 0.14$ T, which arises due to disorder in the bilayer. Inset: two-terminal conductance as a function of density and magnetic field. $B_{off}$ coincides with quantization of the $\nu = \pm 4$ plateaus and the appearance of the $\nu = 0$ state. **c,** $R_{max}(B, T)$ as a function of total applied magnetic field for several angles $\alpha$ between sample and field. Inset: schematic diagram showing the relative orientation between field and sample. **d,** $R_{max}(B, T)$ as a function of the perpendicular component of the magnetic field for various angles. The resistance depends primarily on $B_{perp}$, contradictory to the expected behavior for a Zeeman gap.

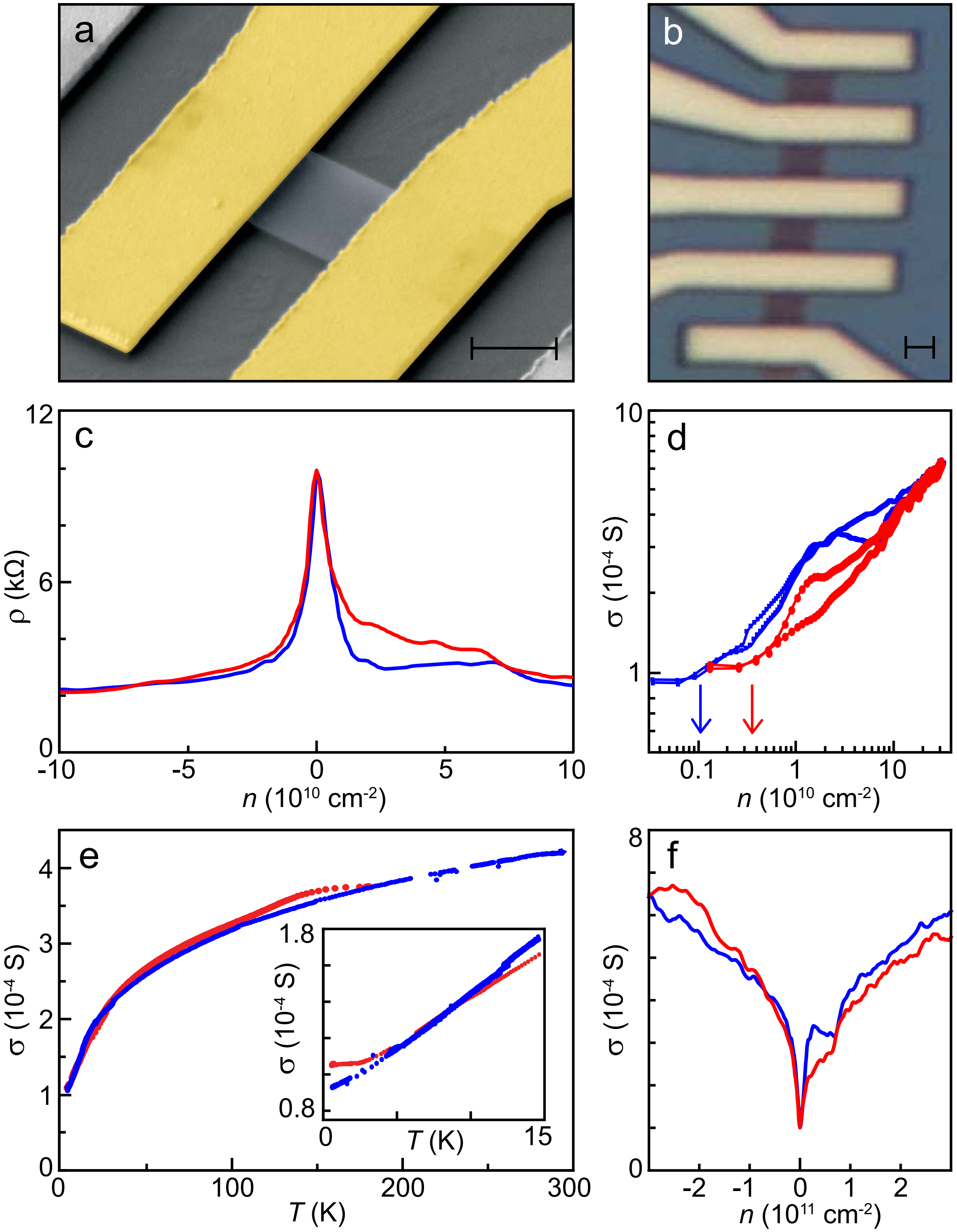

Figure 1

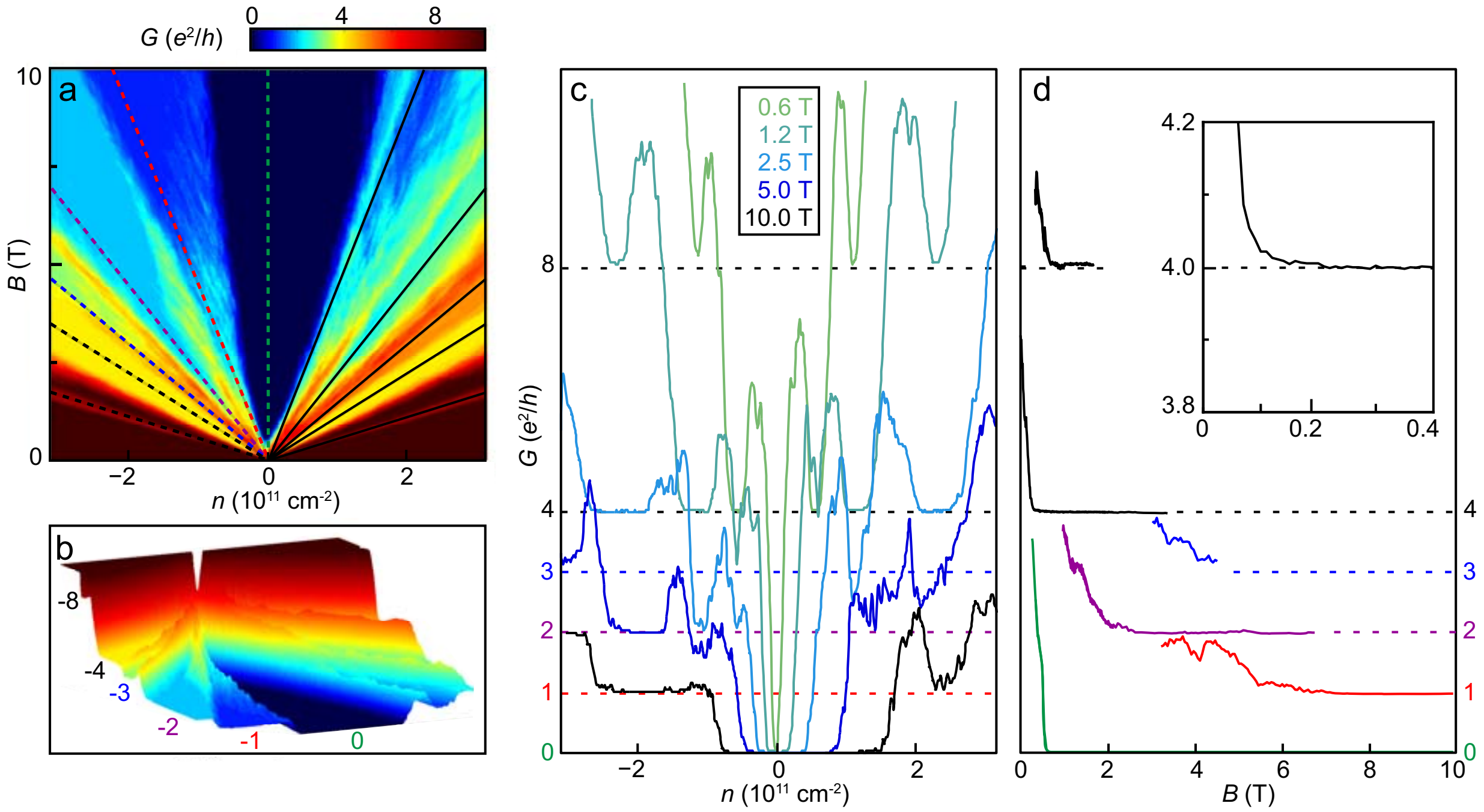

# Figure 2

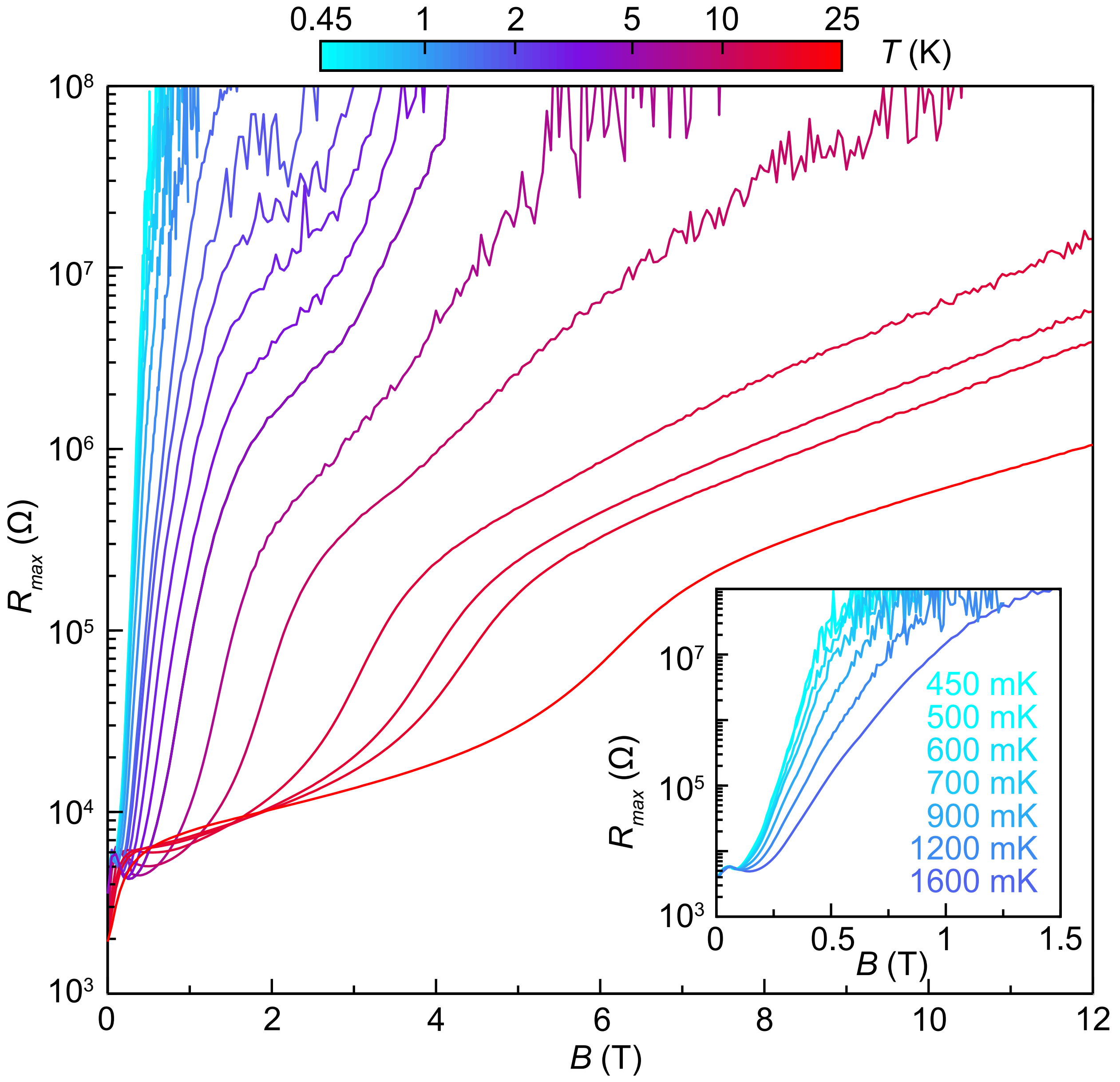

Figure 3

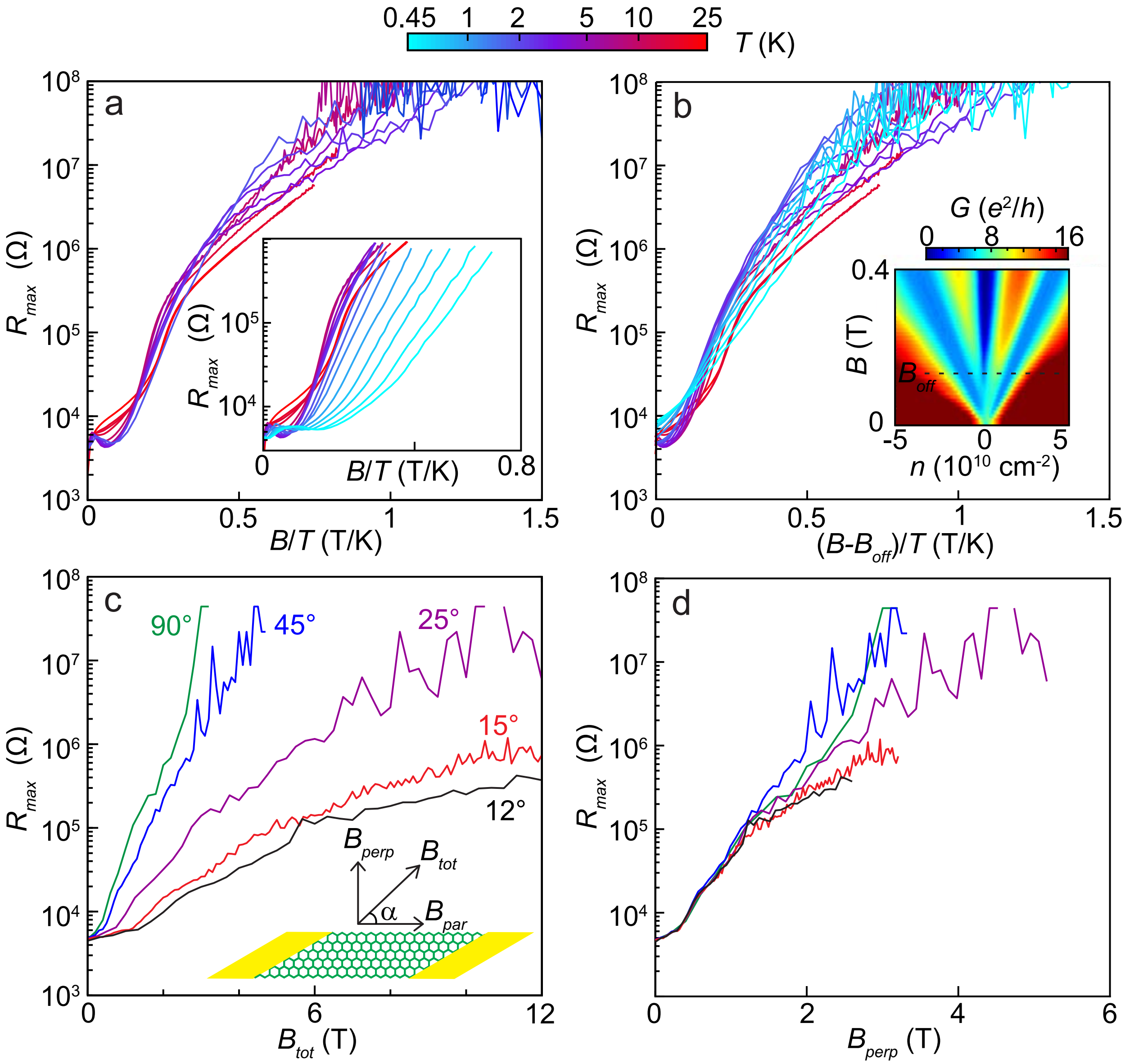
0.45 1 2 5 10 25 T (K)
a
b
c
d
$R_{max}$ (Ω)
B/T (T/K)
(B-B_off)/T (T/K)
G (e²/h)
B (T)
B_off
n (10¹⁰ cm⁻²)
90°
45°
25°
15°
12°
B_perp
B_tot
α
B_par
B_tot (T)
B_perp (T)

Figure 4

**Supplementary Information**

**Resistance in the $\nu = 0$ State**

Fig. S1 shows log[$R(n, B, 4.5\text{ K})$] for sample S3 in a narrow range of carrier density around the charge neutrality point. For $B > 4$ T, $R(n, B, 4.5\text{ K})$ changes by several orders of magnitude as $n$ is varied within the highly resistive region. These oscillations are repeatable, and we find that the regions of relatively low and high resistance are approximately constant in $n$. It is also apparent from Fig. S1 that the highly resistive region is asymmetric: it extends to higher densities for electrons than for holes. Both the oscillations in $R(n, B, 4.5\text{ K})$ and the electron-hole asymmetry have been observed in multiple samples. Their origin is unclear but disorder and/or systematic effects from the fabrication process remain likely candidates. We also note that the position of the peak resistance $R_{max}(B, T)$ shifts slightly (< 50 mV) in back gate voltage $V_{bg}$ as $B$ is varied. We therefore use $R_{max}(B, T)$ rather than the resistance at a constant value of $V_{bg}$ to follow the evolution of the $\nu = 0$ state.

**Hall Bar Devices**

In addition to two-terminal devices, we have also fabricated and measured samples in the Hall bar geometry (Fig. S2a), which exhibit broken symmetry states at $\nu = 0, \pm1$ and $\pm2$. The $\nu = 0$ state displays a large magnetoresistance in four-terminal measurements, so we conclude that this behavior is due to the graphene and is not caused by contact resistance. The longitudinal conductance $\sigma_{xx}$ of sample S5 is plotted in Fig. S2b as a function of $\nu$. Zeros in $\sigma_{xx}$ are clearly visible for $\nu = 0$ and $\nu = \pm4$, and local minima are apparent for $\nu = \pm1$ and $\nu = \pm2$. Fig. S2c shows $\sigma_{xy}$ vs. $\nu$ for the same flake. Again, the $\nu = \pm4$ plateaus are well developed, and a plateau at $\sigma_{xy} = 0$ is also apparent at $\nu = 0$. Quantum Hall plateaus at other intermediate filling factors, however, do not reach their fully quantized values for $B < 6$ T. We therefore conclude that devices in the Hall bar geometry are more disordered than two-terminal devices, likely due to doping from the closely spaced contacts.

The decrease in cleanliness is also evident in the resistance $R$ of sample S5 as a function of magnetic field $B$, which shows a spike at $B \approx 4.5$ T (Fig. S3a). By measuring the two-terminal conductivity between the outer electrical contacts of the Hall bar design while shorting different pairs of inner contacts, we are able to determine that this spike is caused by inhomogeneity between different portions of the flake. Shorting some pairs of inner contacts does not affect the measured resistance, whereas shorting others allows the current to bypass the highly resistive region of the flake so that $R$ at 6 T drops from more than $10^8$ Ω to less than 1 MΩ (Fig. S3b). This not only shows that the flake in inhomogeneous, but also that the contact resistance of the electrical leads (at that of least the two outer contacts) is small compared to the resistances that we measure, providing further evidence that the highly resistive behavior is a fundamental property of the graphene itself.

**Supplementary Information Figure Captions**

**Figure S1 | Fluctuations in resistance in the $\nu$ = 0 state.** Resistance on a log scale as a function of magnetic field and carrier density log[$R$($B$, $n$)] at 4.5 K for sample S3 in a small density range around the charge neutrality point. Oscillations in resistance that are several orders of magnitude are visible as carrier density $n$ is varied. These features occur at constant $n$ as magnetic field is varied.

**Figure S2 | Broken symmetry states in a Hall bar device. a,** Scanning electron micrograph of a typical suspended bilayer graphene flake in the Hall bar geometry. Scale bar is 1 μm. **b,** Longitudinal conductance $\sigma_{xx}$ as a function of filling factor $\nu$ at magnetic field $B$ of 3 T (red), 6 T (green), and 12 T (blue). Zeros are apparent for $\nu$ = 0 and ±4, and local minima occur at $\nu$ = ±1 and ±2. **c,** Hall conductance $\sigma_{xy}$ as a function of $\nu$ at the same magnetic fields as in Figure S2b. The $\nu$ = 0 and ±4 plateaus are fully quantized. Other broken symmetry quantum Hall states at $\nu$ = ±1 and ±2 are apparent, but their Hall conductance is not quantized.

**Figure S3 | Inhomogeneity in Hall bar devices. a,** Four-terminal resistance $R$ as a function of magnetic field $B$ for sample S5. A large spike in $R$ occurs at $B$ = 4.5 T. **b,** Two-terminal resistance of the outer two contacts as a function of $B$ for sample S5 when different pairs of inner contacts are shorted together. Differing behavior as a function of shorted contact indicates sample inhomogeneity. Inset: schematic diagram of the Hall bar device, with a highly resistive region labeled that is consistent with the data. Colors in the main panel correspond to shorting of the pins connected by the same color in the inset.

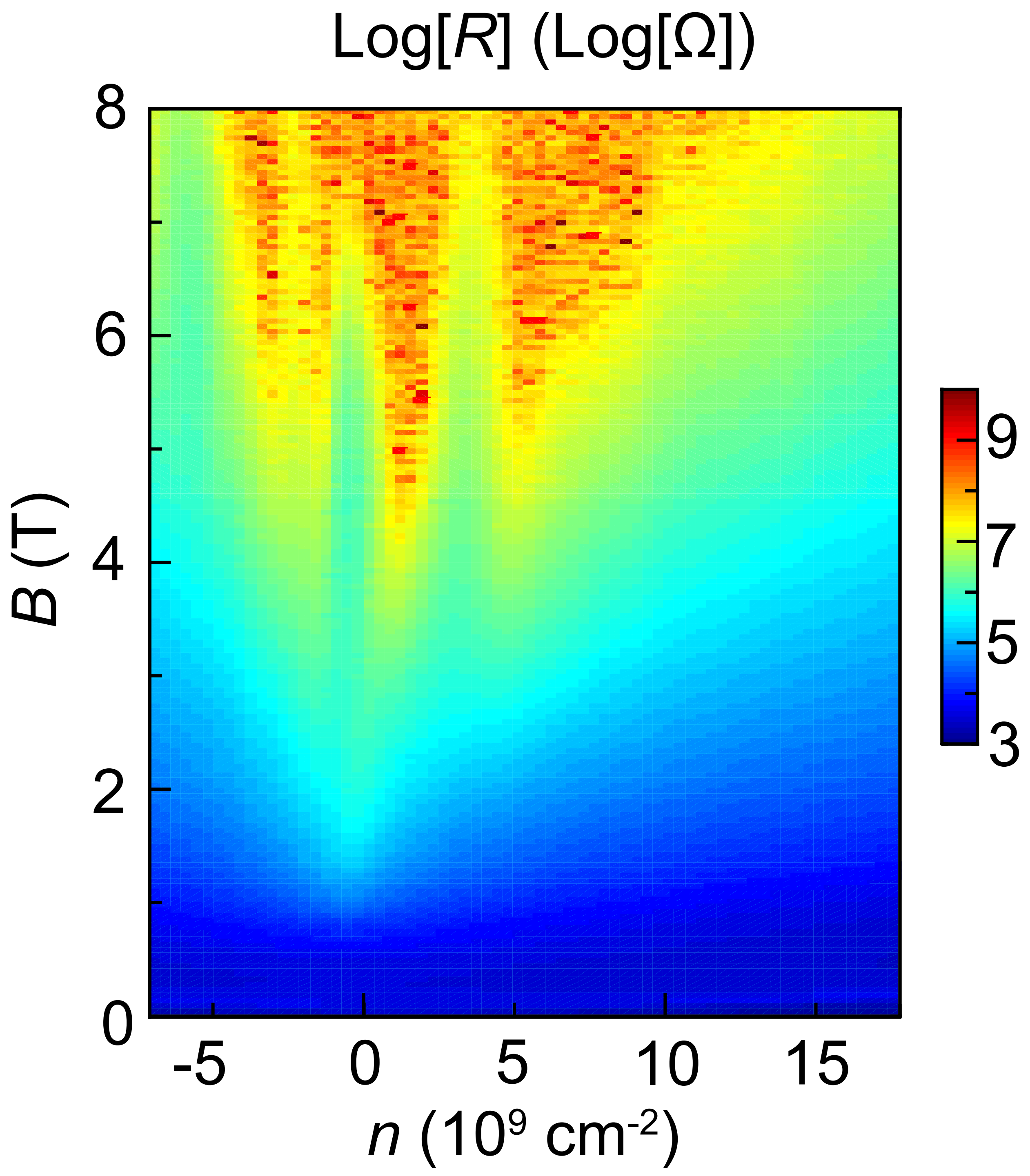

Supplementary Figure 1

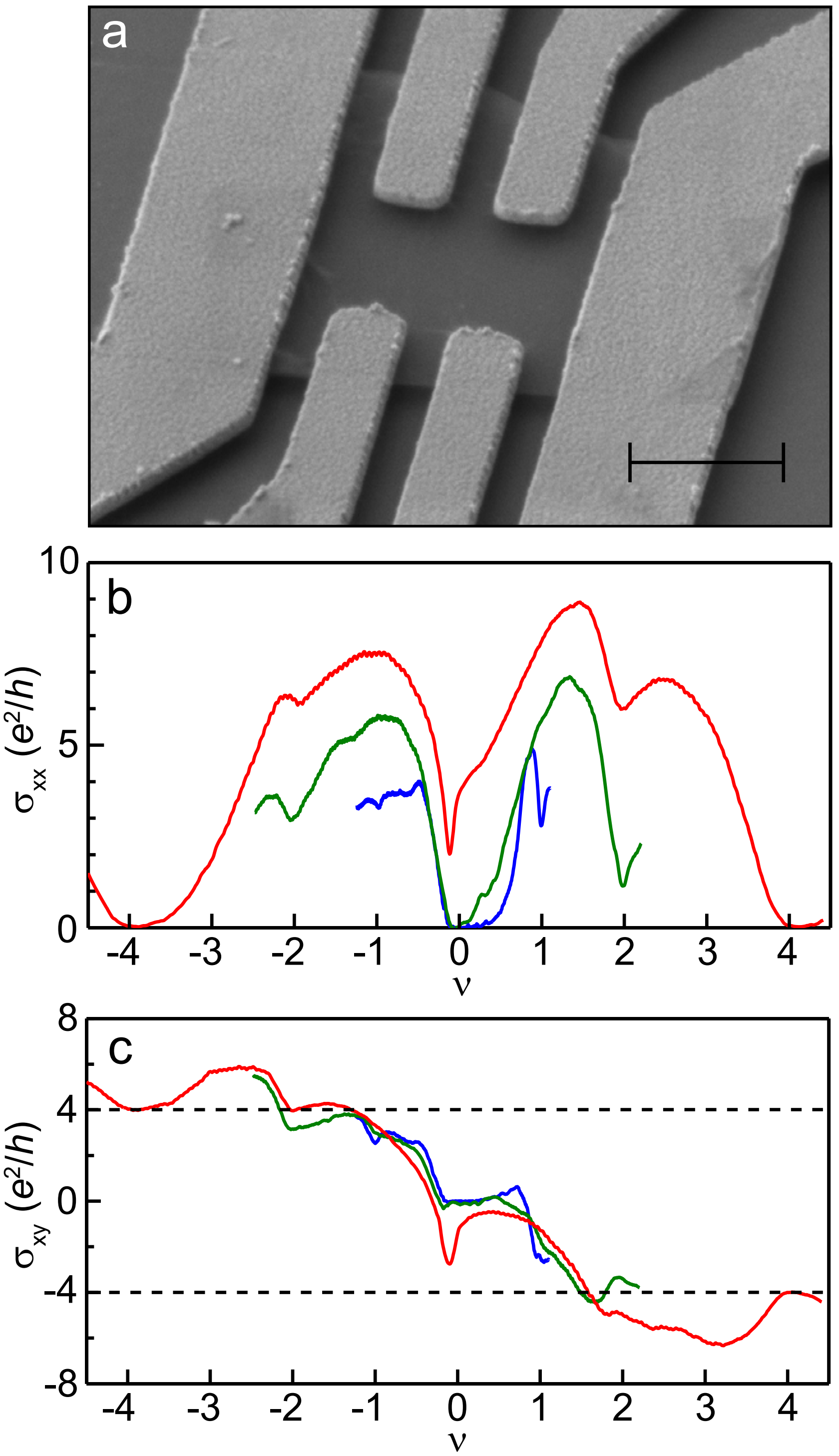

Supplementary Figure 2

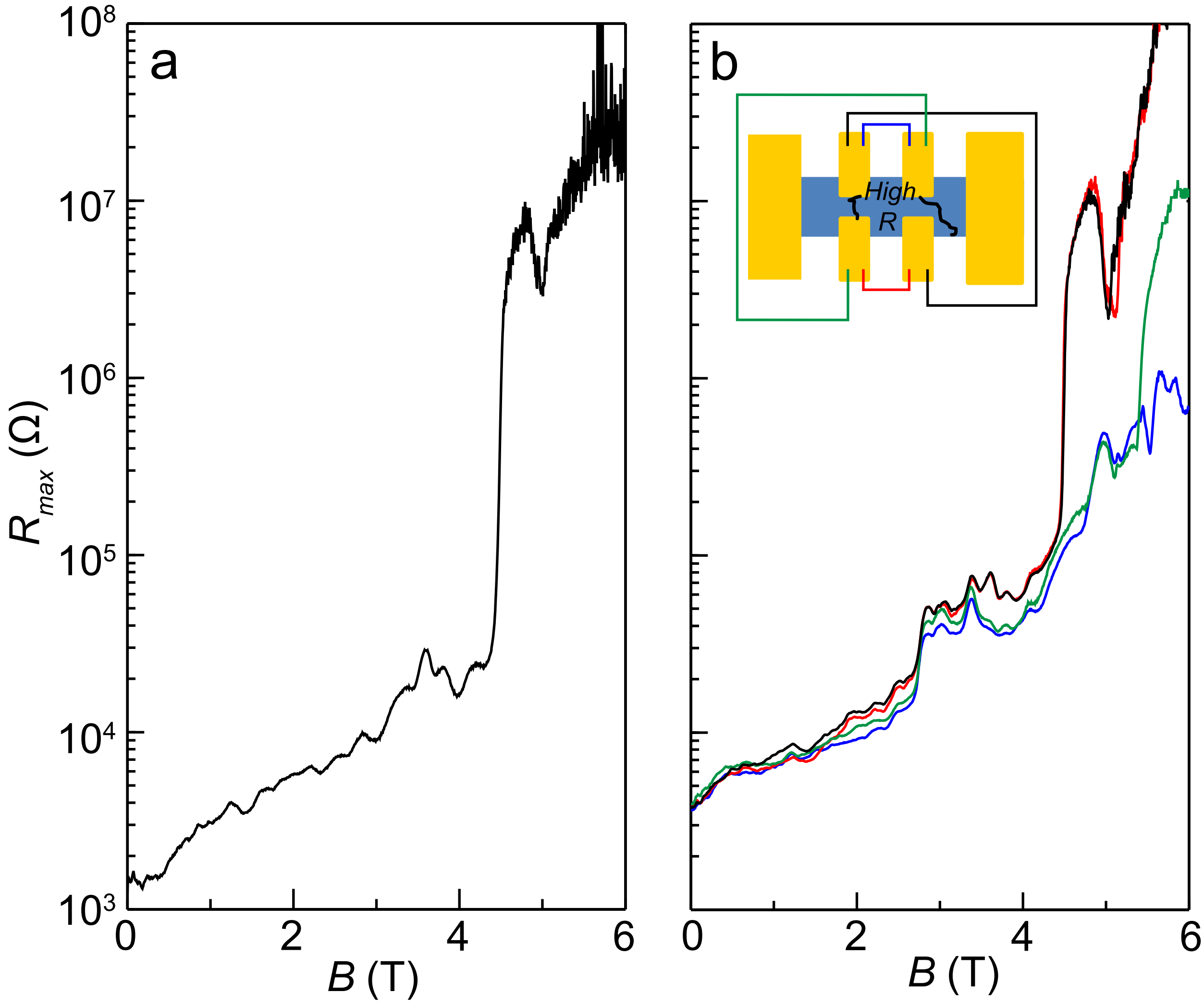

Supplementary Figure 3